\documentclass[aps,prl,
twocolumn,
epsfig]{revtex4}

\usepackage[dvips]{graphics}
\usepackage{graphicx}
\usepackage{amsfonts}
\usepackage{amssymb}
\usepackage{amsmath}

\begin{document}

\title {Landau-Zener Tunneling of Bose-Einstein Condensates in an Optical Lattice}
\author{V.V. Konotop$^1$}
\author{P.G. Kevrekidis$^2$}
\author{M. Salerno$^3$}
\affiliation{
$^1$ Centro de F\'{\i}sica Te\'{o}rica e
Computacional, Universidade de Lisboa, Complexo Interdisciplinar,
Av. Prof. Gama Pinto 2, Lisbon 1649-003, Portugal \\
$^2$  Department of Mathematics and Statistics,
University of Massachusetts, Amherst MA 01003-4515, USA \\
$^3$ Dipartimento di Fisica "E.R. Caianiello",
         Universit\'a di Salerno, I-84081 Baronissi (SA), Italy, and \\
         Istituto Nazionale di Fisica della Materia (INFM), Unit\'a
         di Salerno, Italy
         }

\begin{abstract}
A theory of the non-symmetric Landau-Zener tunneling of Bose-Einstein
condensates in deep
optical lattices is presented. It is shown that periodic exchange
of matter between the bands is described by a set of linearly
coupled nonlinear Schr\"{o}dinger equations. The key role of the
modulational instability in rendering the inter-band transitions
irreversible is highlighted.
\end{abstract}
\maketitle


\paragraph{Introduction.}
After the first realization of Bose-Einstein condensation (BEC) in
an optical lattice (OL)~\cite{AK}, and subsequent experimental
studies of Bloch oscillations and Landau-Zener (LZ) tunneling of
matter~\cite{Morsch1}, numerous theoretical efforts were made to
understand and describe these phenomena. In particular, the effect
of the nonlinearity on LZ tunneling has been recently studied in
Refs. \cite{WN,ZoGa}. Those studies, however, are essentially
based on the assumption of smallness of the lattice potential in
comparison with the recoil energy, which corresponds to a narrow
gap in the spectrum of the underlying linear system -- i.e., to
the case where one can employ a two-mode approximation, where the
modes are plane waves. Meanwhile, recent direct experimental
observation of the LZ tunneling of a BEC in an
OL~\cite{Morsch2,Morsch3} was performed for the range of the
potential amplitudes extended till several recoil energies (a deep
OL), challenging the existing theory~\cite{com}. In the latter
case, the fine structure of Bloch states becomes of crucial
importance and leads to a model different from the one proposed
in~\cite{WN,ZoGa}. Although the basic principles of the LZ
tunneling are hosted in the linear band structure, the nonlinear
interaction makes the phenomenon {\it very different} from the one
known in solid state physics. Indeed, in the presence of
nonlinearity, the  Bloch states can become unstable \cite{KS} and
localized states can appear in the gaps between bands
(gap-solitons) which can dramatically affect the atomic transfer
rate.

The aim of the present paper is to provide a theoretical
understanding of the nonlinear interaction and LZ tunneling
between the lowest bands of a periodic potential. This is
particularly relevant for a deep OL, where the two edges of the
lowest forbidden zone have essentially different properties. This
approach reveals not only the role of modulational instability
(MI) on the atom transfer between zones, but also the effect of
(partial) suppression of the MI by the lattice acceleration. Such
an understanding is important for appreciating the ramifications
of this key linear phenomenon in the presence of nonlinearity and
its connection with critical nonlinear instabilities such as the
MI. It is also particularly relevant for  interpreting recent
experiments in this system \cite{Morsch2,Morsch3}.

\paragraph{Theoretical Approach.} We start with the Gross-Pitaevskii
equation for the macroscopic wave function $\psi=\psi({\bf r},t)$:
\begin{equation}
\label{GPE}
i\hbar \frac{\partial\psi}{\partial t}=-\frac{\hbar^2}
{2m}\Delta\psi+V_{trap}({\bf r})\psi+V_{latt}({\bf r})\psi+g_0|\psi|^2\psi\, ,
\end{equation}
where ${\bf r}=(x,{\bf r}_\bot)$, $g_0=4\pi\hbar^2a_s/m$, $a_s$ is
the $s$-wave scattering length, and $m$ is the atomic mass. The
trap potentials are given by $ V_{trap}({\bf r})=\frac m2
\left(\omega_x^2x^2+ \omega_0^2r_{\bot}^2\right) $ and by $
V_{latt}({\bf r})=\frac{V_0}{2} \cos^2\left[\frac{2\pi}
{\lambda}\left(x-\frac{a}{2}t^2\right)\right] $ where $\omega_x$
and $\omega_0$ are the harmonic oscillator frequencies, $\lambda$
is the laser wavelength and $a$ is the acceleration of the OL.

In the following we restrict to the weakly nonlinear regime which
can be quantified as $\epsilon=(2\sqrt{2}\xi k)^{-1}\ll 1 $ where
$\xi=(8\pi n |a_s|)^{-1/2}$ is the healing length,
$k=\frac{2\pi}{\lambda}$, and $n$ is the mean atomic density. To
this end, we introduce new variables
$X=2k\left(x-\frac{at^2}{2}\right)$, ${\bf R}_\bot=2k{\bf
r}_\bot$, and $T=\frac{8E_{R}}{\hbar}t$, where
$E_{R}=\frac{\hbar^2k^2}{2m}$ is the recoil energy, the
dimensionless parameters $v=\frac{V_0}{16E_{R}}$,
$\gamma=\frac{a\hbar^2k}{(8E_{R})^2}$, and
$\nu_{\alpha}=\frac{\hbar\omega_\alpha}{4E_R}$,
$\sigma=$sign$(a_s)$, and a wave function $\Psi(T,{\bf R})$:
\begin{eqnarray}
\label{wf}
\psi(t,{\bf r})=e^{-ivT+i\gamma XT-i\frac{\gamma^2}{3}
T^3+i\frac{\nu_x^2\gamma^2}{40}T^5}
\sqrt{n} \Psi(T,{\bf R})\,.
\end{eqnarray}
Then, Eq. (\ref{GPE}) can be rewritten in the form
\begin{eqnarray}
\label{GPE_DIM}
    i\partial_T \Psi=\frac 12
    ({\cal L}_X+{\cal L}_\bot)\Psi
    +\gamma X\left(1+\frac{\nu_x^2}{2}T^2\right) \Psi
    \nonumber \\
    +\frac{\nu_x^2}{2} X^2 \Psi
     +\sigma\epsilon^2 |\Psi |^2\Psi
\end{eqnarray}
where ${\cal L}_{X}=-  \partial_X^2  +
v\cos(X)  
$ and $ {\cal L}_{\bot}=-
\Delta_\bot^2+ \nu_0^2R_{\bot}^2. $

For weak nonlinearity the energy of the two-body interaction is
much less than the recoil energy and Eq. (\ref{GPE_DIM}) can be
reduced to a system of coupled one-dimensional nonlinear
Schr\"{o}dinger equations. To show this we introduce a set
of variables $t_l =\epsilon^l T$, $x_l =\epsilon^l X$,
($l=0,1,...$) which are considered independent, and look for a
solution in the form $ \Psi =\psi_1+\epsilon\psi_2+... $, where
$\psi_j=\psi_j(t_0,...,x_0,...,{\bf R}_\bot)$. After substituting
these expansions into (\ref{GPE_DIM}), we single out equations of
the same order in $\epsilon$ as: $ [i\partial_{t_0} -\frac 12
({\cal L}_{x_0}+{\cal L}_{\bot})]\psi_j=F_j $, where $F_1=0$, $
F_2=-i\partial_{t_1} \psi_1,$ etc., and ${\cal L}_{x_0}=
-\partial_{x_0}^2  + v\cos(x_0)$. We then consider the eigenvalue
problems $ {\cal L}_{x_0}\varphi_{\alpha q}(x_0)={\cal
E}_{\alpha}(q) \varphi_{\alpha q}(x_0),$ and ${\cal
L}_{\bot}\zeta_{\mu_\bot}=\mu_{\bot} \zeta_{\mu_\bot}, $ where
$\alpha$ characterizes the zone, $q$ is a wave-vector in the first
Brillouin zone (BZ), and  $ \mu_{\bot}$ refers to the transverse
quantum numbers.

Focusing on the tunneling (i.e., particle transfer) between
the two lowest bands, straightforward algebra shows that if all
atoms are initially concentrated in the lowest transverse state,
the lattice acceleration does not cause (to leading order)
transitions among the transverse levels. This allows us to use
$\zeta_\bot
= (\frac{\nu_0}{\pi})^{1/2}\exp(-\frac{\nu_0}{2}
R_{\bot}^2)$ and look for the
wave function in the
form
\begin{eqnarray}
    \label{gen_sol}
    \psi_1=  e^{-2i \nu_0t_0}\zeta_\bot
    [A_1(x_1,t_2)\varphi_1
            +A_2(x_1,t_2)\varphi_2 ]
\end{eqnarray}
where we denoted with $\varphi_{1,2}$ the modes bordering the BZ
(i.e. the ones with $q=\pi/2$). Further simplification can be
achieved by assuming that the band structure is well pronounced,
i.e., that the condensate is wide enough in the $x$-direction,
$\hbar\omega_x\ll E_{R}$. This is expressed through the rescalings
$\nu_x=\epsilon^2\tilde{\nu}$ and
$\gamma=\epsilon^2\tilde{\gamma}$ where $\tilde{\nu}\sim
\tilde{\gamma} \sim 1$, allowing one to neglect the term
proportional to $\gamma \nu_x^2 T^2$.

We now use the multiple scale technique (see e.g., \cite{KS})
where we take into account the existence of {\em two modes} due to
tunneling. This leads to the system
\begin{eqnarray}
\label{21}
\begin{array}{l}
\displaystyle{
i\partial_{t_2} A_1 + (2m_1)^{-1}
\partial_{x_1}^2 A_1 -\beta_{11}A_1-\beta_{12}A_2
}
\\
\displaystyle{
+(\tilde{\nu}^2x_1^2/2) A_1
-(\chi_{11}|A_1|^2+\chi_{12}|A_2|^2)A_1=0,
}
\\
\displaystyle{
i\partial_{t_2} A_2 + (2m_2)^{-1}
\partial_{x_1}^2 A_2-\beta_{21}A_1-\beta_{22}A_2
}
\\
\displaystyle{
+(\tilde{\nu}^2x_1^2/2) A_2
-(\chi_{21}|A_1|^2+\chi_{22}|A_2|^2)A_2=0.
}
\end{array}
\end{eqnarray}
Here $m_{\alpha}^{-1}=\frac 12 \frac{d^2{\cal
E}_\alpha(q)}{dq^2}$ are the inverse effective masses,
\begin{eqnarray}
\chi_{ij}&=&\frac{\sigma\nu_0}{4\pi^2} \int_{-\pi}^\pi
|\varphi_i(x)|^2|\varphi_j(x)|^2dx \label{nlin}
\\
\beta_{ij}&=& \frac{\tilde{\gamma}}{2\pi} \int_{-\pi}^\pi
x\bar{\varphi}_i(x)\varphi_j(x)dx \label{lin}.
\end{eqnarray}
Recalling that $\varphi_j$ border the lowest gap, one concludes
that i) $m_1<0$ and $m_2>0$; ii) $\beta_{11}=\beta_{22}=0$ and
$\beta_{12}=\beta_{21}=\beta$, and iii) $\chi_{12}=\chi_{21}=\chi$
(below we use the notation $\chi_j=\chi_{jj}$). Then, we end up
with the system
\begin{eqnarray}
\label{finsyst}
i\partial_{t_2} A_\alpha = -(2m_\alpha)^{-1}
\partial_{x_1}^2 A_\alpha+\beta A_{3-\alpha} +(\tilde{\nu}^2x_1^2/2) A_\alpha
\nonumber \\ + (\chi_{\alpha}|A_\alpha|^2+\chi
|A_{3-\alpha}|^2)A_\alpha, \quad \alpha=1,2. \label{21aa}
\end{eqnarray}
Approximating the
Bloch functions
by
$\varphi_\alpha(x)= \sqrt{2}
\cos(\frac{x+(1-\alpha)\pi}{2})$, we find that $\chi
=\frac{\sigma\nu_0}{4\pi}$, $\chi_j=3 \chi$,
$\beta=\gamma$.

At this point, it is worth  to emphasize the differences between
our model and the one used in Refs.~\cite{WN,ZoGa}. First, in a
generic case the nonlinear terms responsible for self-phase and
cross-phase modulation   appear with different coefficients,
respectively $\chi_1$  and  $\chi$, originating from the different
structure of the Bloch functions at two edges of the gap. Unlike
the phenomenological model considered in Ref.~\cite{WN}, these
terms appear in (\ref{finsyst}) with the same sign. Second, our
model includes the group velocity dispersions (i.e. the inverse
effective masses) in explicit form, a feature that does not occur
in the approximation of a shallow OL. Third, the time does not
enter explicitly the equations for the envelopes because it is
scaled out by the ansatz (\ref{wf}) using smallness of
$\gamma\nu_x$, making the resulting system conservative. Finally,
our model takes into account the trap potential (an essential
ingredient of the experimental setup~\cite{Morsch2,Morsch3}).

The system (\ref{21}) leads to a number of
conclusions:

(a) In the presence of nonlinearity the LZ tunneling is an asymmetric
effect. This stems not only from the fact that the coefficients of
the two equations are different i.e., $|m_1|^{-1}\neq m_2^{-1}$
and $\chi_{1}\neq\chi_2$, but also form the fact that the
condensate at one gap edge is modulationally stable [edge 2 (1) in
the case of a positive (negative) scattering length] and
modulationally unstable at the other edge~\cite{KS}.

(b) If, however, one considers the evolution of un-modulated Bloch
waves (i.e., the case $\partial_{x_1} A_j=0$) then the inverse
effective masses do not appear in the description of the
phenomenon and in a rather common situation when $\chi_1\approx
\chi_2$, the LZ tunneling becomes symmetric. This case
 can be explicitly treated {\it analytically}.
The analysis of the resulting equations for the populations of the
two bands demonstrates the existence of oscillations of matter
between the bands. However, as this case ignores the important
feature of MI, we do not discuss it here.

(c) Although  Eqs. (\ref{21}) are coupled for
$\gamma=0$, the number of particles ${\cal N}_j=\int |A_j|^2dx_1$,
is separately preserved  (i.e. $d{\cal N}_j/dt_2=0$ if $\gamma=0$)
and LZ tunneling cannot occur in this case. On the contrary, for
$\gamma\neq 0$ one has the conservation of the total number of
particles ${\cal N}={\cal N}_1+{\cal N}_2$: $d{\cal N}/dt_2=0$ and
LZ tunneling can occur.

\paragraph{Numerical Results.}

We define the rate characterizing tunneling as $r(t)={\cal N}_1/{\cal
N}$. In the numerical experiments, we have simulated the 2-state
transition model of Eqs. (\ref{21aa}) with parameters adapted to
the  experiments reported in Ref.~\cite{Morsch2}. In particular,
we consider $V_0=2.2 E_{R}\approx \hbar\times 10^3$Hz,
corresponding to $v=0.1375$; hence, $m_1^{-1}\approx -1.5$ and
$m_2^{-1}\approx 2.5$ (as obtained from the band structure of the
respective potential). A weak magnetic trapping for rubidium atoms has been
implemented by choosing $\omega_x\approx 2\pi\times 55Hz$ and
$\omega_0\approx 2\pi\times 200$Hz, which corresponds to $\nu_x=
0.03$ and  $\nu_0= 0.1$. The nonlinear and linear
couplings have been obtained from Eqs. (\ref{nlin})-(\ref{lin}).
In particular, $\chi=0.008$ and $\chi_j=0.024$,
while $\beta$, similarly to the setting of \cite{Morsch3}, was varied
(typically) in
the interval $\beta\in [0.2,1.4]$. Furthermore, the calculations
were also performed for solutions of various amplitudes
to emulate the variation of the nonlinearity
in \cite{Morsch3}.

\begin{figure}[t]
\includegraphics[width=8.cm,height=4.26cm,angle=0,clip]{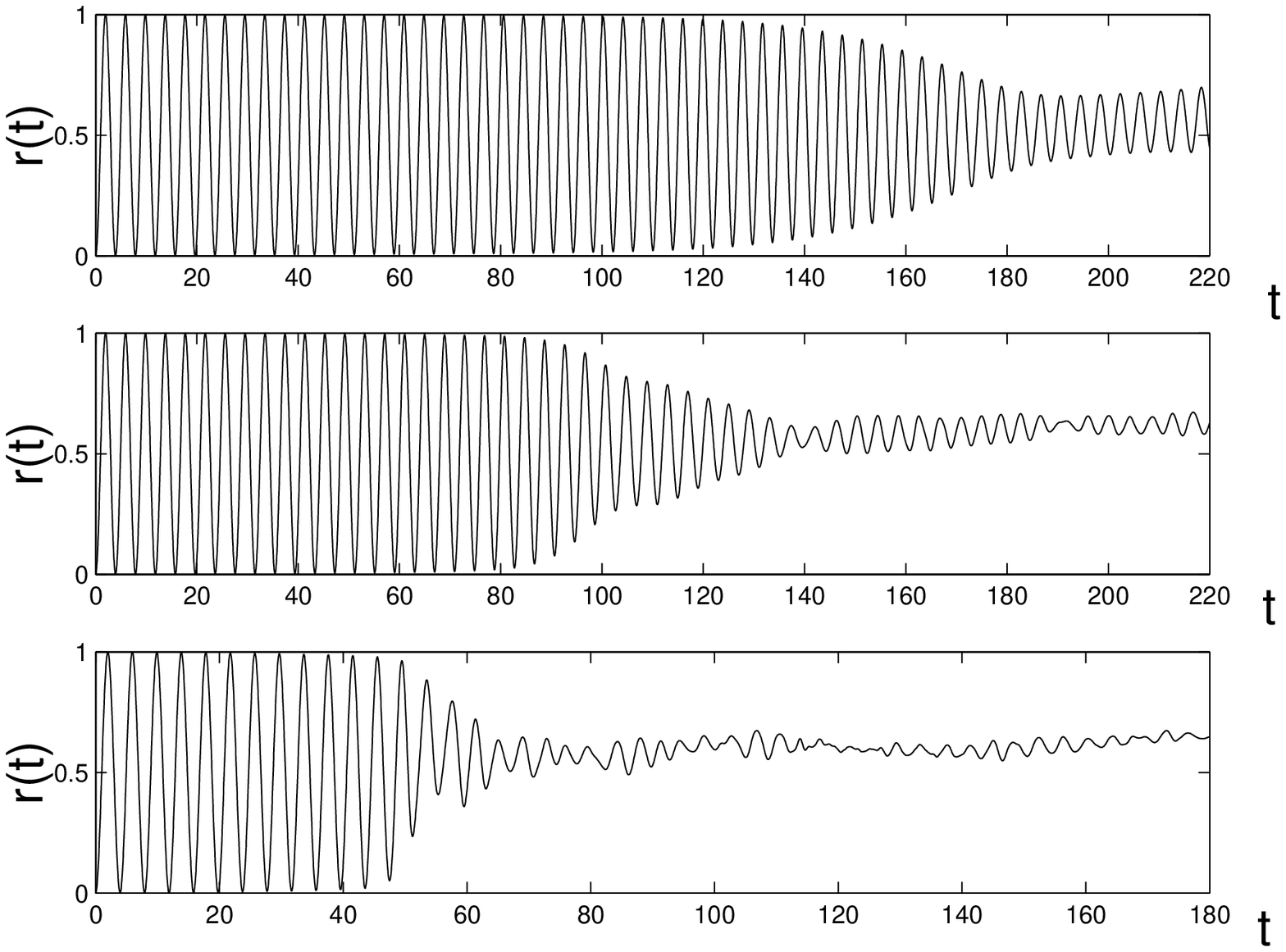}
\includegraphics[width=8.cm,height=4.26cm,angle=0,clip]{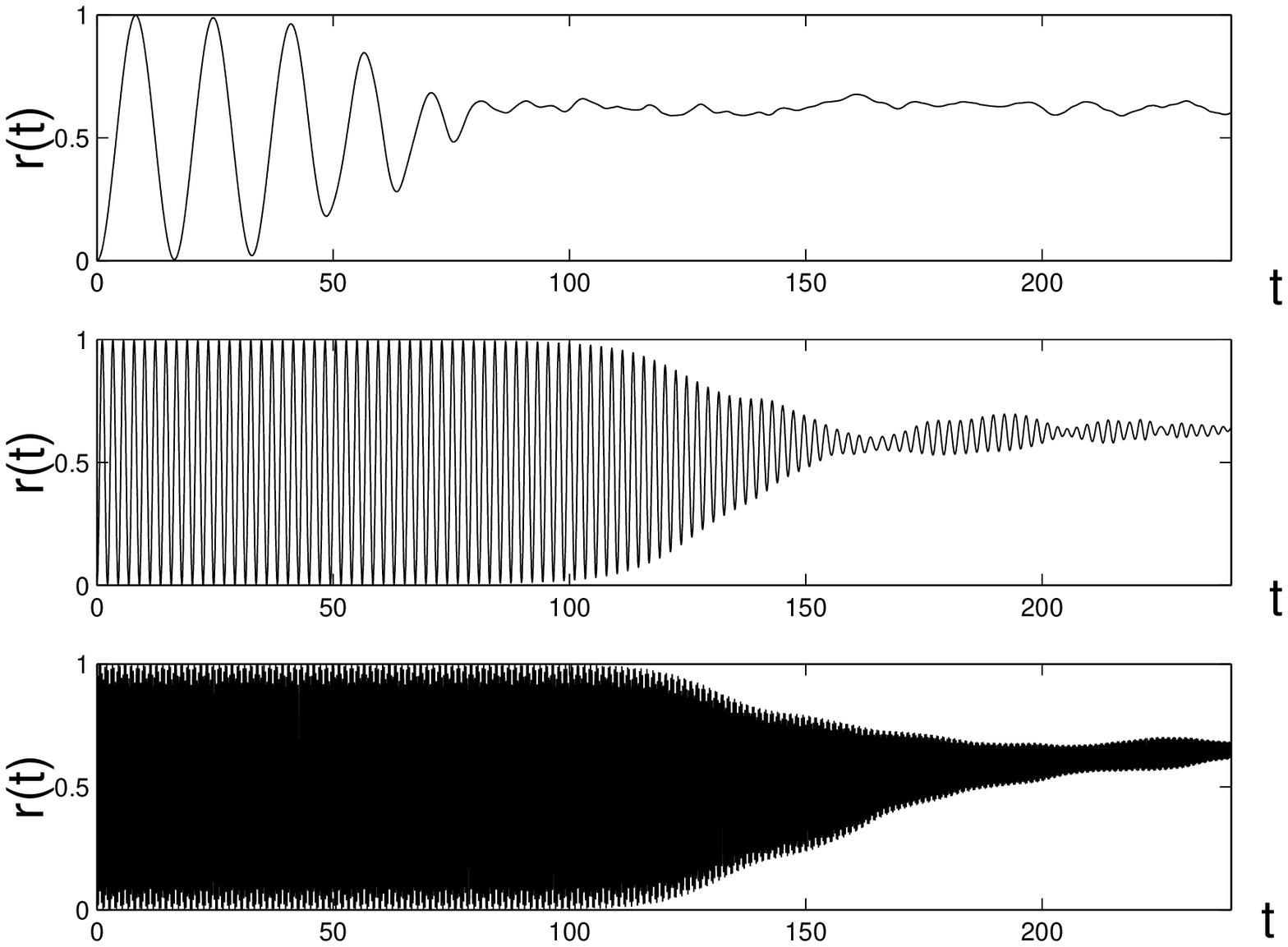}
\caption{The tunneling ratio $r(t)=N_1/N$ is given for three
different amplitudes and $\beta=0.8$ on the top three panels:
$\max(A_2(x,0))=3.22$ (first from the top), $4.83$ (second) and
$6.46$ (third). It is also shown for $\max(A_2(x,0))=4.83$, for
three different couplings $\beta=0.2$ (fourth panel), $1.4$ (fifth)
and $10$ (sixth).} \label{Fig1}
\end{figure}

We first examined cases where $A_1(x,0)=0$ and the second
component is an exact solution of Eq. (\ref{21aa}) with $\alpha=2$ and
in the absence of linear coupling. Typical results for the transfer rate are
shown in Fig. \ref{Fig1}.
\begin{figure}[t]
\includegraphics[width=8.cm,height=4.26cm,angle=0,clip]{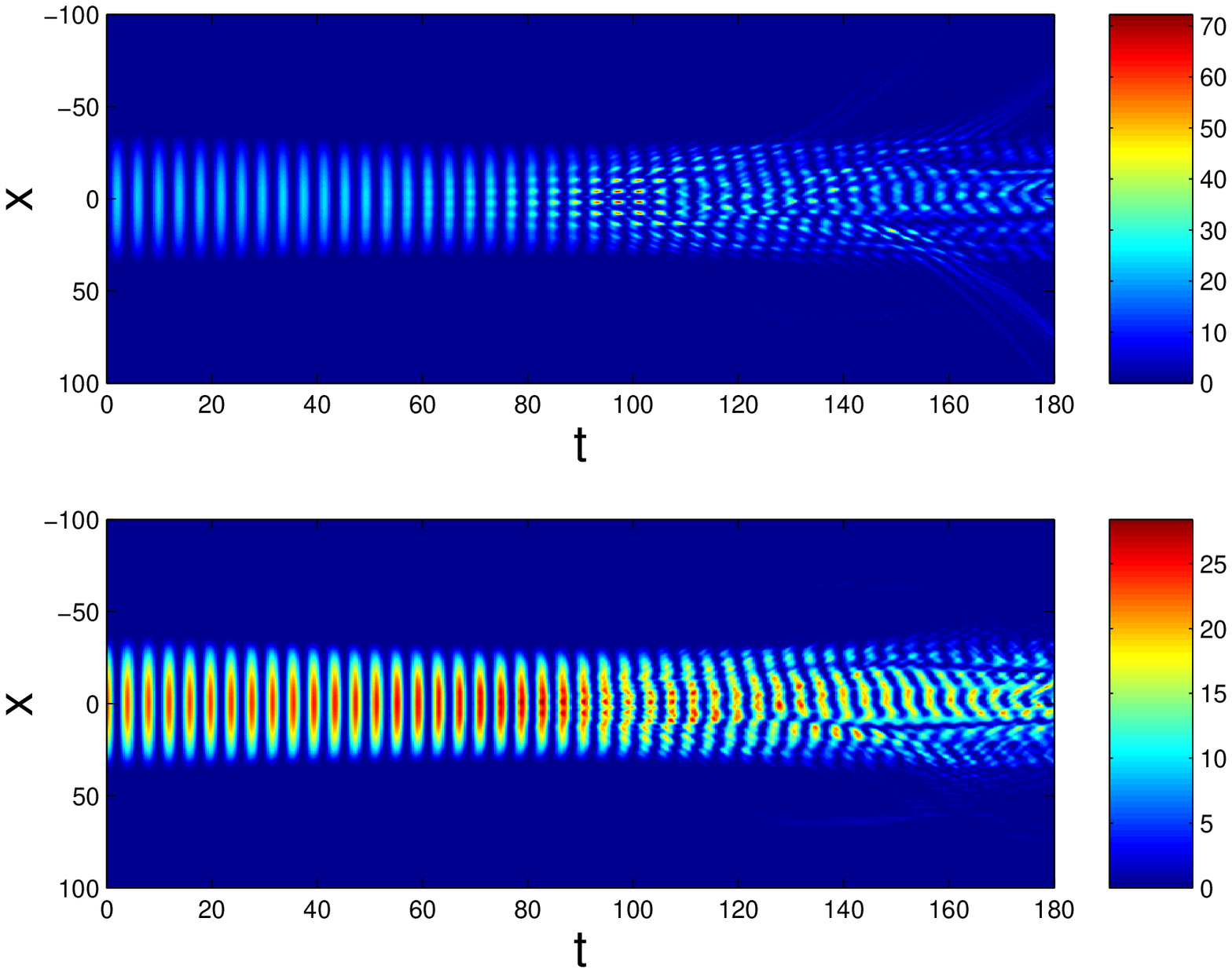}
\includegraphics[width=8.cm,height=4.26cm,angle=0,clip]{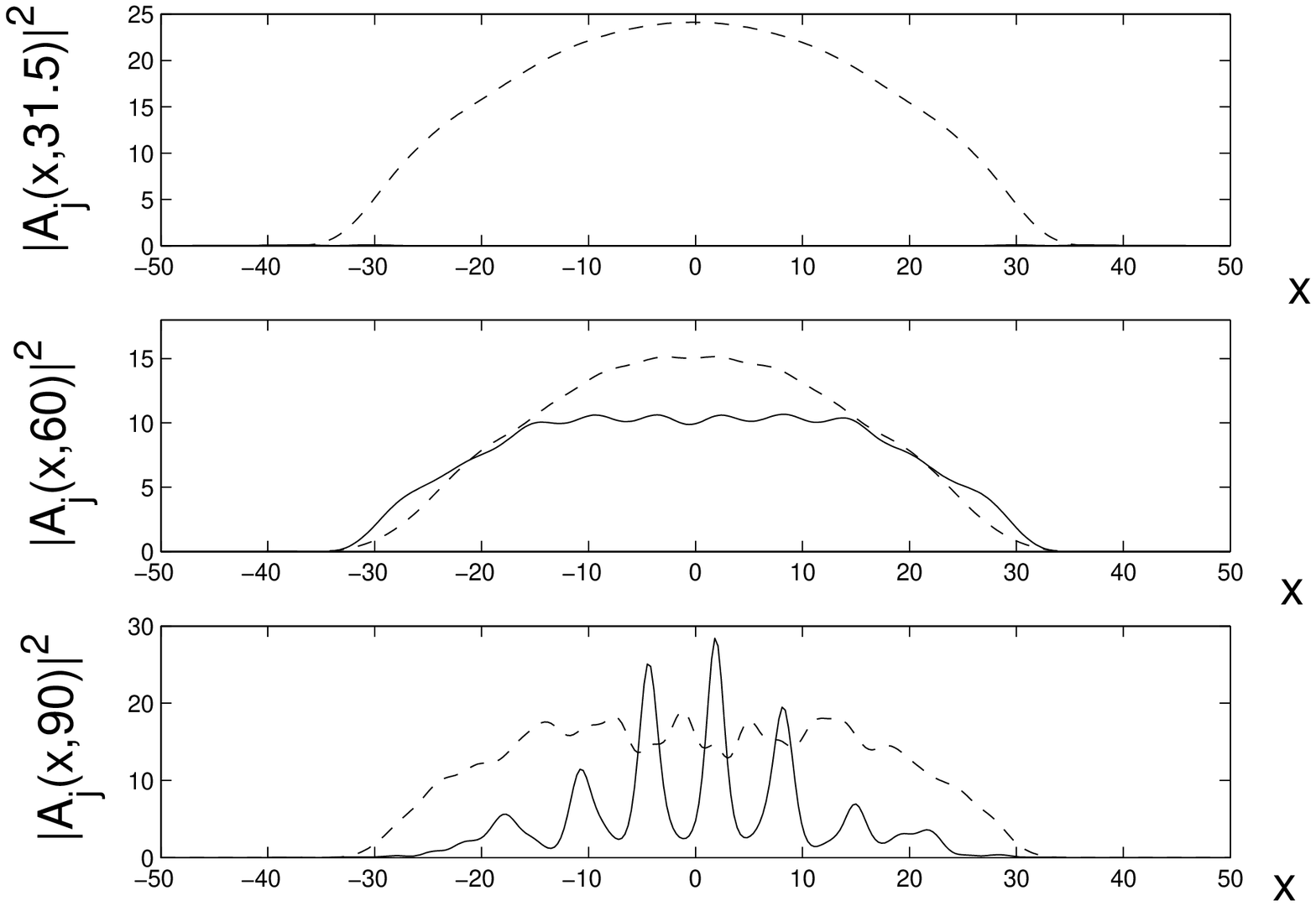}
\caption{The first two panels from the top show the
spatio-temporal contour plots of $|A_1(x,t)|^2$ (first from the
top) and $|A_2(x,t)|^2$ (second panel). Initially (i.e., at $t=0$)
the second
component is populated [$\max(A_2(x,0))=4.83$], while the first is
not. The remaining three panels show three snapshots at different
times of the spatial profile of the first (solid line) and of the
second (dashed line) component. In the middle and bottom  panels the
development of the MI is evident.}
\label{Fig2}
\end{figure}
At the early stages of the dynamical evolution, one observes
periodic particle transfer between two bands, reflecting the fact
that the probabilities of direct and reciprocal tunneling are the
same (an effect not taken into account by the simplified
qualitative picture in \cite{Morsch3} but also predicted by the
model of Ref.~\cite{ZoGa}). However, after some critical time,
$t_{cr}$, the behavior of the system changes drastically: the
amplitude of the tunneling rate oscillations decreases
substantially, indicating that the system never returns back to
the state where all atoms are condensed in only one band.
As one can see from the top three  panels in Fig.~\ref{Fig1}, the
critical time is smaller for larger initial atomic populations of
the second zone i.e., for larger nonlinearity. The remaining
panels of Fig. \ref{Fig1} display two other important features:
(i) the frequency of oscillations increases with the lattice
acceleration and (ii) the critical time also increases with the
lattice acceleration. The first of these features seems to be
intuitively clear: larger linear force corresponds to larger
probability of tunneling and thus to more intensive interchange of
atoms between the bands. The second item, as well as the existence
of $t_{cr}$ itself, are less evident. In order to appreciate these
features, we computed the evolution of the density profiles
corresponding to the two bands. Fig. \ref{Fig2} shows the
spatio-temporal contour plot (as well as snapshots of time)
evolution of the two components for an initial condition with
$\max(A_2(x,0))=4.83$. Approximately at $t=35$ (for the case
$\epsilon^2\sim 0.1$, corresponding in physical units to
approximately 90 $\mu$s), MI starts to develop in the system and
at $t=60$ one can already see a well pronounced periodic structure
of the distribution of atoms in the first band,  which at $t=90$
develops into  a number of solitary pulses, as is typical for MI.
The emergence of stable localized states eventually renders the
transfer of matter between the two zones irreversible. The time of
the development of soliton-like structures coincides with the
critical time observed in Fig. \ref{Fig1} (second top panel). This
explains the change in the dynamics of the tunneling rate reported
above. Namely, the particle exchange between the two bands is
suppressed after solitonic (train) structures are developed from
the ``unstable'' band (the lowest band in our case). Notice that
the stronger the nonlinearity is, the larger is the region of
unstable wavenumbers and equivalently the shorter is the time for
the instability development, thus explaining the differences in
the tunneling rates shown in  Fig.~\ref{Fig1}. One can also
explain the dependence of $r(t)$ on the lattice acceleration
(linear force). Indeed, while in the ``unstable'' band, the
nonlinearity tends to fragment the condensate, immediately after
transfer to the ``stable'' band the nonlinearity leads to
dispersion of the wave packet. By increasing the linear force
($\beta$) one increases the frequency of oscillations of $r(t)$,
i.e., decreases the time atoms spend in the ``unstable'' band,
thus effectively reducing the MI growth rate. This explains the
increase of the critical time $t_{cr}$ at which MI develops,
observed for increasing $\beta$ in the three bottom panels of Fig.
\ref{Fig1}; notice from the bottom panel that for very large
$\beta$ the effect saturates. In all the cases we observed rather
weak dependence of the final upper-to-lower tunneling on the
nonlinearity which corroborates the results of
Ref.~\cite{Morsch3}. Our result of the tunneling rate however is
larger than the experimentally observed, which, apparently, is an
artifact of the two-mode model; the latter does not account for
the atoms tunneling to higher bands, thus reducing the effective
number of atoms participating in inter-band transfer with respect
to the experiment.

The case in which all atoms are initially in the lowest band can
be treated similarly. Since, in this case, the homogeneous
distributions are unstable, it is natural to start with an initial
condition which is a gap soliton resulting from MI and containing
the same number of atoms as in the previous case. To understand
the phenomenon, we use as initial condition a soliton which is
somewhat displaced from the bottom of the potential  so that the
tunneling matter will be spatially separated from the pulse (due
the dynamics in the trap) and therefore easily distinguishable.
This setting is illustrated in Fig. 3. As is shown in the snapshot
of the bottom right panel of the figure, the wave packets on the
left are well separated from the localized pulse. These represent
the matter that has tunneled to the second band. Our numerical
results indicate that to each burst of matter in the second band,
there corresponds one in the first band. Also notice that the
burst of matter in the second band (see the top right panel of
Fig. 3) is becoming more and more extended as one expects for a
Bloch wave of the stable band. The separation of the matter
strongly decreases the interaction between the bands (the
corresponding wave packets have practically zero overlapping),
thus leading to the formation of a plateau in the tunneling
rate\footnote{On a longer time scale the dynamics in the trap give
rise to a more complicated behavior of the tunneling rate with the
occurrence of a plateau each time the matter in the two bands gets
separated by the dynamics.}. The asymptotic tunneling rate
achieved in this case may be different from the one obtained in
previous case (cf., e.g., Fig.~\ref{Fig1}); this further supports
the existence of an essential asymmetry induced by the
nonlinearity in the LZ tunneling \footnote{We should emphasize
importance of the trap potential for understanding the experiments
of \cite{Morsch3}: due to the  turning points, the zero-velocity
atoms monitored in the experiment may belong to both bands.}.
\begin{figure}[t]
\includegraphics[width=8.4cm,height=4.75cm,angle=0,clip]{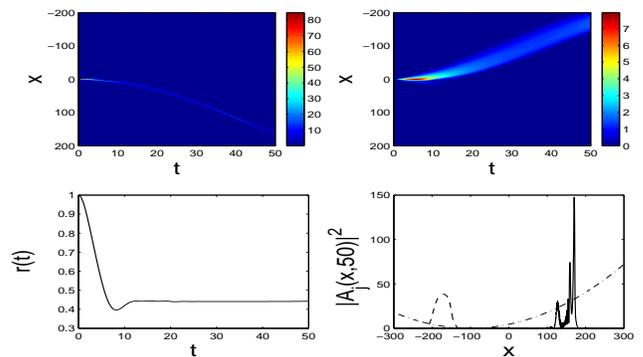}
\caption{The top left and right panels show space-time contour
plots of the 1st and 2nd band wavefunctions respectively. The
bottom left shows the temporal evolution of the tunneling ratio
$r(t)$, while the bottom right shows a snapshot of the solution at
$t=50$. Solid line: $|A_1|^2$; dashed line: $|A_2|^2$;
dash-dotted: parabolic potential.} \label{Fig3}
\end{figure}
\paragraph{Conclusions.} We have provided a systematic
derivation of a two-band model reduction relevant to the
interaction of the lowest bands of a periodic potential. A theme
of particular interest, in this context, is the Landau-Zener
tunneling of a BEC between two bands of an optical lattice, a
recurring theme of theoretical and experimental importance. While
a number of our results, like the sensitivity of the effect with
respect to initial distributions, the oscillatory behavior of
tunneling rate, and the qualitative symmetry with respect to
exchange of bands $1\leftrightarrow2$ with simultaneous change of
the sign of the scattering length corroborate the predictions of
\cite{ZoGa}, we have illustrated important differences from
previously explored models and have highlighted the asymmetric
nature of the tunneling and the important role of the nonlinearity
through the modulational instability in the process. These results
illustrate the importance of (experimentally) studying this system
for longer times substantially larger than the critical time
(and/or stronger nonlinearities) to appreciate the role of the
nonlinearity in modifying the classical picture of the
Landau-Zener tunneling.

VVK is grateful to O. Morsch for stimulating discussions.
PGK acknowledges support from NSF-DMS-0204585, NSF-CAREER and the
Eppley Foundation for Research. MS acknowledges financial support
from a MURST-PRIN-2003 Initiative.


\begin{thebibliography}{99}
\bibitem{com}
Experimental upper-to-lower band tunneling rates are lower than
the ones for the reciprocal tunneling (see, Figs. 2b and 3 of
\cite{Morsch3}). This does not agree with the theoretical
predictions, stating the opposite (see Fig. 2a of \cite{Morsch3}).
\end{thebibliography}
\end{document}